\documentclass{vgtc}                          % final (conference style)
\graphicspath{{figures/}{pictures/}{images/}{./}} % where to search for the images

\usepackage{times}                     % we use Times as the main font
\usepackage{tabu}                      % only used for the table example
\usepackage{booktabs}                  % only used for the table example
\usepackage{lipsum}                    % used to generate placeholder text
\usepackage{mwe}                       % used to generate placeholder figures

\usepackage{mathptmx}                  % use matching math font

\usepackage{tabularx}
\usepackage{multirow}
\usepackage{algorithm}
\usepackage{algpseudocode}
\usepackage{textcomp, gensymb} 
\usepackage{amssymb}                  % use matching math font
\usepackage{amsmath}
\usepackage{xurl}
\usepackage[autostyle, english = american]{csquotes}
\MakeOuterQuote{"}
\usepackage[table,svgnames,dvipsnames]{xcolor}
\newcommand{\highlighttextbf}[1]{{#1}}
\newcommand{\highlightcell}[1]{\cellcolor{LightGreen!50}\highlighttextbf{#1}}
\newcommand{\greencell}[1]{\cellcolor{LightGreen!50}#1}
\newcommand{\yellowcell}[1]{\cellcolor{Yellow!50}#1}
\newcommand{\redcell}[1]{\cellcolor{Red!50}#1}

\newcommand\shl[1]{%
  \bgroup
  \hskip0pt\color{cyan!80!green}%
  #1%
  \egroup
}

\vgtccategory{Research}

\vgtcinsertpkg

\title{The Perceptual Cost of Passthrough: How Video See-Through HMDs Degrade Human Visual Perception of Acuity, Contrast, and Color}

\author{
Jialin Wang$^{1}$,
Songming Ping$^{2}$,
Kemu Xu$^{2}$,
Yue Li$^{2}$,
Hai-Ning Liang$^{1,3}$%
\thanks{Corresponding author. E-mail: hainingliang@hkust-gz.edu.cn}
}

\affiliation{\scriptsize
$^{1}$Computational Media and Arts Thrust, The Hong Kong University of Science and Technology (Guangzhou), Guangzhou, China\\
$^{2}$Academy of Artificial Intelligence and Advanced Technology, Xi'an Jiaotong-Liverpool University, Suzhou, China\\
$^{3}$Division of Arts and Machine Creativity, The Hong Kong University of Science and Technology, HKSAR, China
}

\teaser{
  \centering
  \includegraphics[width=\linewidth]{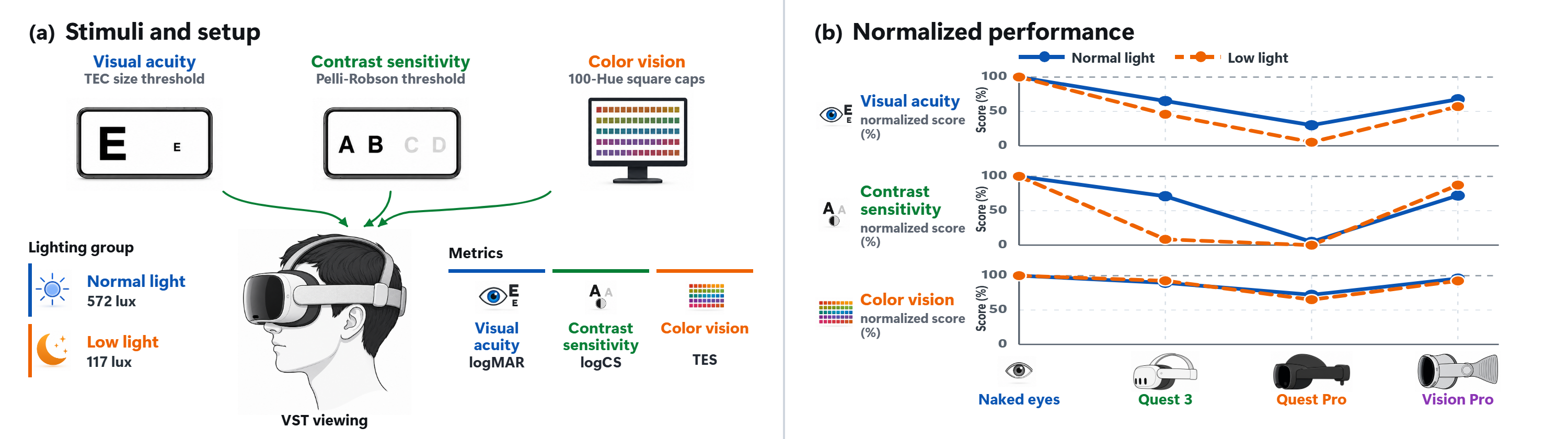}
  \caption{Overview of the VST visual-performance benchmark. (a) Stimuli, viewing setup, lighting groups, and metrics. (b) Median normalized scores relative to each participant's naked-eye baseline after global min-max normalization; logMAR and TES were inverted so higher scores indicate better performance.}
  \label{fig:workflow}
}

\abstract{
Video see-through (VST) technology aims to seamlessly blend the virtual and physical worlds by reconstructing reality through cameras. However, while manufacturers promise high perceptual fidelity, it remains unclear how closely recent commercial VST systems preserve basic visual functions across environmental conditions. In this work, we present an end-to-end perceptual benchmark for three popular VST headsets: Apple Vision Pro, Meta Quest 3, and Meta Quest Pro. Using adapted psychophysical measures, we evaluated participants' visual acuity, contrast sensitivity, and color vision under both normal and low-light conditions, with naked-eye vision as the reference. Our results show measurable gaps between VST and naked-eye performance, especially for visual acuity and contrast sensitivity in low-light environments. By mapping these perceptual gaps across devices, visual functions, and lighting levels, this work provides a practical benchmark for current commercial VST capabilities and highlights where experience design or device optimization may need to compensate for perceptual loss.
} % end of abstract


\begin{document}


\firstsection{Introduction}

\maketitle

\begin{table*}[!ht]
\centering
\begingroup
\setlength{\tabcolsep}{3pt}
\begin{tabular}{p{1.5cm}<{\raggedright}*{1}{p{2.4cm}<{\centering}}*{4}{p{2.1cm}<{\centering}}*{2}{p{1.4cm}<{\centering}}}
\toprule
HMD & Arch. & Cam. res. & Cam. pix. & Cam. ap. & Ctr. PPD & Dis. col. depth & Dis. dyn. range\\
\midrule
Quest Pro &  \yellowcell{single RGB + two depth cam.} & \yellowcell{4608 $\times$ 3456 + 1280 $\times$ 1024}& \yellowcell{16 + 1.3 megapixels} & \redcell{Unknown} & \yellowcell{$\sim$30} & \redcell{Unknown} & \greencell{SDR}\\
\midrule
Quest 3 & \greencell{Two RGB cam.} & \redcell{Unknown}& \redcell{Unknown} & \redcell{Unknown} & \greencell{18} & \redcell{Unknown} & \greencell{SDR}\\
\midrule
Vision Pro & \greencell{Two RGB cam.}& \redcell{Unknown}& \greencell{6.5 megapixels} & \greencell{18 mm ƒ/2.00} & \yellowcell{$\sim$55} & \redcell{Unknown} & \greencell{HDR}\\
\bottomrule
\end{tabular}
\endgroup
\caption{Comparison of key performance factors for three major VST HMDs with known information. Factors from official and third-party sources are highlighted in \colorbox{LightGreen!50}{green} and \colorbox{Yellow!50}{yellow} colors. \colorbox{Red!50}{Red} indicates factors missing from both official and third-party sources. Cam., res., pix., ap., Ctr., Dis., col., and dyn. denote camera, resolution, pixels, aperture, central, display, color, and dynamic, respectively.}
\label{tab:keyFactors}
\end{table*}

Passthrough refers to techniques that aim to address the limitations of Virtual Reality (VR) by leveraging outward-facing cameras to reconstruct images that users would otherwise see without the VR head-mounted display (HMD) \cite{NeuralPassthrough}---that is, with their naked eyes. Passthrough is a narrower concept than VST and is primarily used for VR. To enhance clarity and reduce ambiguity, passthrough can be defined as an application of VST in VR HMDs. VR HMDs utilizing passthrough can also be classified as VST HMDs or mixed reality (MR) HMDs. 

VST HMDs represent a significant evolution in immersive technology, enabling users to blend virtual and real-world environments seamlessly. Current VST HMDs work by transmitting real-world images into the VR space captured via external cameras, creating an MR experience within the VR framework. This technology combines the high immersion level of VR with the real-world view of augmented reality (AR), resulting in an MR experience different from optical see-through (OST) HMDs such as the Hololens 1/2, Meta 2, and Magic Leap \cite{steffen2019framework}. This hybrid approach has found applications in various fields, including video games, training simulations, remote collaboration, and surgical operations \cite{orione2024utilization,Armstrong2024,Dhawan2024, 10765408}, and offers other innovative interactions where users benefit from interacting with both digital and physical elements simultaneously \cite{guo2024breaking, PassthroughPiano, rhee2020augmented, wang2022realitylens, 10.1145/3772318.3790315, 10937409}.

In \Cref{tab:keyFactors}, we compare publicly reported camera- and display-side factors for Meta Quest 3, Meta Quest Pro, and Apple Vision Pro using official specifications and third-party reports\footnote{\url{https://support.apple.com/en-us/117810} \url{https://www.meta.com/quest/quest-pro/tech-specs/} \url{https://www.meta.com/quest/quest-3/} \url{https://developers.meta.com/horizon/design/display/} \url{http://wellsenn.com/} \url{https://sadlyinreality.com/the-final-meta-quest-pro-analysis/} \url{https://kguttag.com/2023/08/09/apple-vision-pro-part-5b-more-on-monitor-replacement-is-ridiculous/}}. Architecture, resolution/pixel count, aperture, and central PPD summarize camera arrangement, captured detail, light capture, and central VST sharpness. The two display-side columns report color depth and dynamic range only when disclosed. Because documentation is incomplete and third-party estimates may vary, these entries support qualitative interpretation and should not be read as end-to-end passthrough measurements.

Due to limited documentation from both primary and secondary sources, many specifications remain undisclosed, making it challenging to distinguish among the different VST HMDs. This limitation underscores the difficulty of understanding VST performance from device specifications alone. End-to-end perceptual testing is therefore needed to evaluate what users can actually see through the full VST pipeline, including cameras, image processing, displays, calibration, and environmental lighting. The ultimate goal of VST HMDs is to achieve visual fidelity that matches or closely approximates normal human vision. This level of fidelity is essential for applications where precise visual perception is critical, such as in medical training, teleoperation, or navigating complex environments \cite{ruthenbeck2015virtual, Watanabe2020, ihemedu2017virtual, AdaptiveVoice}. Several key visual perception abilities, including visual acuity, contrast sensitivity, and color vision, are crucial and can be measured using a low-cost psychophysical benchmarking method based on standard stimulus displays and simple calibration equipment. Measuring them provides valuable insights into how well these devices replicate natural human vision.

By quantifying these visual perception metrics, both manufacturers and users can identify specific areas where VST HMDs fall short and implement targeted compensatory strategies. For instance, understanding the limitations in visual perception can inform the development of adaptive contrast enhancement techniques or the strategic reduction of passthrough usage in scenarios where visual acuity is compromised. These compensations can significantly mitigate the negative impact of poor visual performance, leading to more comfortable and effective user interactions. Recent research has leveraged traditional visual metrics and display-based visual tests to drive visual perception standards for HMDs, notably through adapting visual acuity charts into VR environments like the Omnidirectional Virtual Visual Acuity (OVVA) \cite{ovva} and establishing the ultimate resolution limits for sharp vision \cite{ashraf2025resolution}. These approaches show the feasibility of adapting vision-science methods to immersive displays, but they do not provide a unified perceptual benchmark of recent commercial VST devices across visual acuity, contrast sensitivity, color discrimination, naked-eye reference viewing, and different lighting conditions.

The main goal of this work is to provide an accessible end-to-end benchmark for measuring the visual perception of VST HMDs. In short, our contributions include:
\begin{itemize}
\item Development of an accessible framework for measuring and comparing VST HMDs using three standard visual functions: visual acuity, contrast sensitivity, and color vision (see \Cref{sec:metrics} and \Cref{sec:programs}). The source code and supplementary material are available at \url{https://github.com/Chaosikaros/VST-Visual-Perception-Benchmark}.
\item A detailed evaluation of visual acuity, contrast sensitivity, and color vision through controlled experiments and user studies (see \Cref{sec:conditions} to \Cref{sec:results}).
\item Identification of the limitations of current VST HMDs and areas needing improvement (see \Cref{sec:discussion}).
\item Insights that can improve future developments of VST HMDs and inform users about their effectiveness for everyday applications, particularly in implementing compensatory design strategies to mitigate perceptual shortcomings (see \Cref{sec:insights}).
\end{itemize}
By comparing recent commercial VST HMDs against naked-eye viewing across these basic visual functions, this work aims to clarify where current devices preserve perceptual performance and where experience design compensations may be needed.

\section{Related Work}
\subsection{Visual Acuity: Its Importance and Measurements}
Visual acuity is a fundamental aspect of visual perception and is commonly referred to as clarity or sharpness of vision. Normal human visual acuity is essential for tasks that require fine detail recognition. For VST HMDs, achieving normal visual acuity ensures that users can seamlessly transition between virtual and real-world elements without losing important visual information.
Low visual acuity in VST HMDs can result from several factors, including the resolution of the cameras used, the quality of the lenses, and the display resolution \cite{PassthroughPlus}. Distortions from fish-eye lenses and insufficient pixel density can also degrade image clarity \cite{Pohl2013}.

In addition to hardware factors, software issues such as poor image processing algorithms, low render resolution, inadequate calibration between the virtual and real-world views, compression artifacts, and suboptimal image fusion can further reduce the visual acuity \cite{NeuralPassthrough, Carkeet2021, ovva}.
Although VST HMDs are not designed for outdoor scenarios, some users have attempted to walk or drive their cars while wearing a VST HMD, either seeking novelty of experience or fun, which can be risky in complex environments and fast-moving scenarios. Studies have focused on the safety issues associated with walking while wearing VST HMDs \cite{sousa2019safe}. Furthermore, vision science research has indicated that low visual acuity is related to a fear of falling and a cautious walking strategy \cite{KLEIN2003644, HALLEMANS2010547}. In addition, a study showed that most patients, who had diseases causing low visual acuity and who were currently driving, had at least 20/40 visual acuity (0.5 in decimal format) in the better-seeing eye \cite{PATNAIK2019336}. Having 20/40 visual acuity is also one of the visual acuity standards for meeting the minimum drivers' licensing requirements, which aim to distinguish safe drivers from unsafe ones \cite{wood2005standard}.

The concept of a retinal display aims to promote a display that matches normal human visual acuity. However, most VR HMDs still cannot replace a normal monitor due to a low PPD of around 20 compared to the retinal display standard of around 60 PPD, which affects text quality \cite{lu2023display}. PPD is also important for assessing the visual clarity of VST. However, just like with VR HMDs, many manufacturers do not publish official PPD values for VST. Typically, only those with higher performance provide these metrics, such as the Meta Quest 3, which has 18 PPD, and the Varjo XR-4 and Varjo XR-4 focal edition with 33 PPD and 51 PPD, respectively. It is worth noting that these values only represent part of the central PPD, that is, the PPD of the middle or focal area, since PPD is often highest at the center and gradually decreases towards the edges \footnote{\url{https://www.meta.com/quest/quest-3/} \url{https://varjo.com/products/xr-4/}}. 

However, VST HMDs can rely more on end-to-end metrics like visual acuity. People's visual acuity has traditionally been measured using standardized vision charts, such as the Snellen chart \cite{hussain2006changing}. In VR, researchers have adapted these methods to measure the visual clarity of images \cite{ovva}. Vision charts can measure the combined effect of camera resolution, lens quality, display resolution, render resolution, and image processing on end-to-end visual acuity. Prior work on advanced visual acuity charts has also shown that charts with a continuous measuring range can detect subtle differences that traditional discrete charts (e.g., 0.1, 0.2 to 1.0) may miss \cite{10.1007/978-3-030-89029-2_44, ovva}. Building on this line of work, we use an enhanced visual acuity chart as one component of a broader VST benchmark that compares multiple commercial devices with naked-eye viewing under different lighting conditions (see \Cref{sec:programs}).

\subsection{Contrast Sensitivity: Its Importance and Measurements}
Contrast sensitivity is the ability to detect luminance differences between objects and their background, supporting the perception of outlines, patterns, and textures of very small objects. It plays a significant role in low-light levels and environments with varying lighting \cite{rahimi2021image}. 

In VST HMDs, good contrast sensitivity ensures that users can perceive fine visual details in complex environments. Notably, better contrast sensitivity is more important than better visual acuity for tasks such as driving \cite{wood2005standard}. The quality of the cameras' sensors, the dynamic range, and the performance of the HMD's display affect contrast sensitivity \cite{Skorka2011}. The cameras used in VST HMDs share similar limitations with those in mobile phones due to their small apertures, which restrict the number of photons they can gather, leading to noisy images in low-light conditions \cite{Hasinoff2016}. Such cameras with poor low-light performance can fail to capture necessary details, compromising the user experience as the low-light environments are also common for VR users \cite{wueller2013low, liba2019handheld, luidolt2020gaze}. While image processing algorithms that can enhance contrast, reduce noise, and manage exposure are important for cameras, poorly optimized algorithms can lead to washed-out images or excessive noise, reducing contrast sensitivity \cite{ren2018lecarm, ying2017new}. However, VST HMDs typically use the same algorithms for all light levels since they are not specifically designed as vision-enhancement systems \cite{11474752}. 

Contrast sensitivity is traditionally measured using sine-wave gratings and contrast sensitivity charts (e.g., the Pelli-Robson chart). These methods are well-established for assessing human visual performance in clinical and research settings. Sine-wave gratings involve patterns of alternating light and dark bars with varying spatial frequencies and contrast levels, allowing for a detailed analysis of contrast sensitivity across different scales \cite{haughom2013sine}. The Pelli-Robson chart, on the other hand, presents letters at a fixed size but with decreasing contrast, providing a straightforward assessment of overall contrast sensitivity \cite{Pelli1988}. Moreover, monitor-based contrast sensitivity charts have emerged as a convenient alternative to traditional methods \cite{Kollbaum2014}. These digital charts can be easily adjusted to test a wide range of contrast levels and spatial frequencies, making them a flexible tool for contrast sensitivity assessment. In this work, we adapt this measurement logic to VST by designing a digital Pelli-Robson chart for both VST HMDs and naked-eye viewing (see \Cref{sec:programs}). This contrast-sensitivity measure complements visual acuity and color discrimination, allowing the benchmark to compare several basic visual functions rather than a single display property.

\subsection{Color Perception: Its Importance and Measurements}
Color vision is the ability to distinguish different hues, which is essential for tasks that rely on color coding and aesthetic judgment. Accurate color representation in VST HMDs ensures that users can correctly interpret real-world signals and maintain a natural visual experience. However, unlike traditional OST HMDs like the HoloLens, both the camera sensor and HMD display panel in VST HMDs can involve color distortion. The color accuracy of the cameras and the display's color gamut are critical factors. Cameras with poor color fidelity or displays that cannot reproduce a wide range of colors will result in inaccurate color representation and even misrepresentation. Color correction algorithms, white balance adjustments, and calibration between the camera feed and the display are also essential to mitigate these issues. Inadequate software processing can lead to color distortions and mismatches, affecting user performance and color perception \cite{finlayson2015color, nam2010color}.

Furthermore, poor color accuracy is harmful to specific careers or tasks that rely on accurate color vision. For example, MR-based smart manufacturing for factories is an important scenario where VST HMDs can be applied \cite{juraschek2018mixed}. Electricians in such factories often work with color-coded wiring systems, and the ability to distinguish between different wires is crucial to prevent dangerous mistakes \cite{grzybowski2019color}. Similarly, normal color vision is also a requirement for obtaining a driver's license, as drivers need to interpret color-coded signals, indicators, and navigational aids to ensure safe driving operations \cite{johnson2005vision}. Beyond these practical applications, professionals such as graphic designers, artists, photographers, videographers, and fashion designers rely heavily on accurate color representation and perception. These individuals often use high-quality monitors to ensure that their work is represented correctly, and any color distortion in VST HMDs could significantly impact their ability to perform their tasks effectively \cite{ovva, NeuralPassthrough}. As such, these individuals are unlikely to adopt VR or MR systems that do not preserve reliable color appearance. 

Color vision has traditionally been assessed using tests like the Ishihara test and the Farnsworth-Munsell 100 hue test (the 100-Hue test for short) \cite{tamura2017light, Farnsworth1943}. The Ishihara test is primarily used to detect red-green color deficiencies, while the 100-Hue test measures the ability to discern small differences in hue, providing a detailed profile of color vision across the spectrum. In VR, researchers have adapted the 100-Hue test to evaluate the color accuracy of the virtual environments in VR systems \cite{Cwierz2021}. De Souza and Tartz used a simplified 100-Hue test on several VST HMDs \cite{de2024visual}. Our study extends this direction by using a digital 100-Hue task as part of a multi-metric benchmark, rather than treating color discrimination as an isolated device characteristic. Moreover, a digital version of the 100-Hue test is as feasible as the physical version in common scenarios \cite{Murphy2015}. The available digital 100-Hue implementation was Flash-based and is no longer practical to deploy. Therefore, we developed a Unity implementation for this study. As VST technology continues to advance, the importance of accurate color perception and representation is only going to grow further, making ongoing research and development in this area crucial for the future of immersive experiences.

\subsection{Limitations of Video See-through on Environmental Adaptation}
Environmental adaptation is a critical aspect of visual perception, including the ability to adjust to varying lighting conditions rapidly and effectively \cite{Kwon2019}. This capability is essential for maintaining clear and consistent visual information across different environments. For ideal VST HMDs, effective environmental adaptation ensures that users can seamlessly perceive and interact with both virtual and real-world elements, regardless of changes in lighting conditions. However, one of the primary limitations of current VST HMDs is their struggle with adapting to rapidly changing lighting conditions. Existing HMD-vision studies have motivated perceptual testing for immersive displays \cite{ovva, de2024visual}, but they have not yet provided a unified VST benchmark that combines multiple basic visual functions with explicit lighting-level comparisons across recent commercial devices.

Unlike the human eye, which can quickly adjust to variations in light intensity through mechanisms such as pupillary light reflex (constriction in brighter conditions and dilation in dimmer conditions) and photoreceptor adaptation (the eye's ability to adjust its sensitivity to light, becoming more sensitive in the dark and less sensitive in bright light), cameras in VST HMDs often fail to keep up with these changes \cite{mathot2018pupillometry, gross2008human}. This can result in images that are either overexposed or underexposed, leading to loss of detail and reduced visual clarity \cite{ovva}. The cameras used in VST HMDs are typically constrained by their sensor technology and structure. Furthermore, the image processing algorithms used to enhance visual output can sometimes introduce artifacts or fail to adequately adjust to changes in lighting, further degrading the visual experience \cite{NeuralPassthrough}. 

Overall, the limitations of VST technology in environmental adaptation are caused by the inherent differences between camera and display technology and the human visual system. The human eye's rapid adaptation to varying lighting conditions is the result of complex biological processes that current VST HMDs cannot fully replicate. Taken together, prior work establishes the relevance of visual acuity, contrast sensitivity, and color discrimination for immersive displays, while our work contributes a unified end-to-end comparison of these functions for VST HMDs against naked-eye viewing.

\section{Visual Perception Tests for VST}

Based on the importance of visual perception and the lack of related metrics, we present a low-cost psychophysical benchmarking method that adapts vision-science tests to VST. Our method consists of 3 digital vision tests for visual acuity, contrast sensitivity, and color vision.

\subsection{Metric Definitions} \label{sec:metrics}
\subsubsection{Visual Acuity}
Visual acuity is a key measure of human vision, defined based on the visual angle. It measures the angle that an object subtends at the eye, describing the apparent size of an object as seen by the observer. It is calculated using \Cref{eq:visualAngle},  with \textit{Object size} representing the actual size of the object being observed and \textit{Object distance} describing the distance from the observer's eye to the object. 
\begin{equation} \label{eq:visualAngle}
\text{Visual Angle in degrees} = 2 \cdot arctan(\frac{\frac{\text{Object Size}} {2}} {\text{Object Distance}})\\
\end{equation}

The Snellen fraction is a traditional way to represent visual acuity, such as 20/20 vision, which refers to normal visual acuity, allowing a person to see what an average individual can see on an eye chart from 20 feet (or about 6 meters) away. In this notation, the numerator refers to the distance at which the test is performed (usually 20 feet/6 meters), and the denominator refers to the distance at which a person with normal vision can read the same line on the chart. For example, 20/40 vision means a person can see the chart as clearly at 20 feet away as someone with “normal” vision would see it from 40 feet away. Another related metric is the minimum angle of resolution (MAR), which is the smallest gap size that can be resolved by the eye, measured in arc minutes (1 arc minute is 1/60 of a degree). Decimal visual acuity is the reciprocal of MAR, which also provides a straightforward representation of the Snellen fraction. 

logMAR (logarithm of the MAR) is another way to quantify visual acuity. logMAR is often used in ophthalmic research because it provides a more uniform scale distribution than decimal visual acuity. However, decimal visual acuity and Snellen fraction are more commonly used in clinical settings due to their straightforward and intuitive nature. Normal visual acuity is typically defined as 20/20 in Snellen format, 1.0 in decimal format, and 0 in logMAR \cite{holladay2004visual}. For Snellen and decimal formats, higher numbers indicate better visual acuity. Conversely, in the logMAR scale, values greater than 0 indicate worse than normal vision, and values less than 0 indicate better than normal vision. The relationship among visual angle, MAR, Snellen fraction, decimal visual acuity, and logMAR is summarized in \Cref{eq:visualAcuity}.  
\begin{equation} \label{eq:visualAcuity}
\begin{split}
&\text{MAR} = \text{Visual Angle (VA) in arc min} = \text{VA in degrees} \times 60\\
&\text{Decimal VA} = \frac{1} {\text{MAR}} = \frac{\text{Numerator of Snellen fraction}}{\text{Denominator of Snellen fraction}}\\
&\text{logMAR} = \log_{10}(\text{MAR}) = -\log_{10}(\text{Decimal VA})
\end{split}
\end{equation}

\subsubsection{Contrast Sensitivity}
Contrast sensitivity is another important measurement and is described as the ability to discern differences in luminance between an object and its background. It is particularly crucial for tasks such as night driving, where the ability to detect low-contrast objects is essential. However, contrast is often assessed using different metrics depending on the nature of the stimuli.
Weber contrast is commonly used for stimuli that are on a uniform background, such as letters or other simple shapes. It is defined as \Cref{eq:WeberContrast}.
 \begin{equation} \label{eq:WeberContrast}
\begin{split}
&\text{Weber Contrast} = \frac{L_{\text{target}} - L_{\text{background}}}{L_{\text{background}}} \\
&L_{\text{target}} \text{ is the luminance of the target}\\
& L_{\text{background}} \text{ is the luminance of the background}
\end{split}
\end{equation}
For periodic patterns such as sine wave gratings, the contrast of an image can be quantified using the Michelson contrast formula. 
Root mean square (RMS) contrast is used for more complex, natural images and is calculated as the standard deviation of pixel intensities. 
Each type of contrast metric is preferred for different types of stimuli: Weber contrast for letter stimuli, Michelson contrast for gratings, and RMS contrast for natural stimuli and efficiency calculations \cite{pelli2013measuring}. In this work, we focus on Weber contrast due to its relevance for letter stimuli on the Pelli-Robson chart.
 
Contrast sensitivity is defined as the inverse of the contrast threshold, which is the minimum contrast level at which an observer can detect a pattern. Log contrast sensitivity (logCS) is simply the logarithm of the contrast sensitivity. This measure is often used because it can better represent the wide range of contrast sensitivities that the human visual system can perceive.
Contrast Percentage Threshold is the minimum contrast level, expressed as a percentage, at which an observer can detect a pattern. These relationships are expressed as \Cref{eq:ContrastSensitivity}.
\begin{equation} \label{eq:ContrastSensitivity}
\begin{split}
&\text{Contrast Sensitivity} = \frac{1}{\text{Contrast Threshold (CT)}}\\
&\text{Log Contrast Sensitivity} = \log_{10}\left(\frac{1}{\text{CT}}\right)\\
&\text{Contrast Percentage Threshold} = \text{CT} \times 100\%
\end{split}
\end{equation}
A higher contrast sensitivity indicates that an observer can detect lower contrast levels, which in turn corresponds to better visual performance. Normal contrast sensitivity varies depending on the spatial frequency of the stimulus (i.e., how fine or coarse the pattern is). For the Pelli-Robson chart, a logCS of 2.0 represents normal contrast sensitivity at 1\% contrast threshold. (or a contrast sensitivity score of 100). A Pelli-Robson contrast sensitivity score of less than 1.5 is indicative of visual impairment, while a score of less than 1.0 signifies visual disability \cite{mantyjarvi2001normal}.

\subsubsection{Color Vision}
Color vision is another critical aspect of visual performance, assessing the ability to distinguish between different colors. Various metrics and tests are used to evaluate color vision, with one of the most prominent being the Farnsworth-Munsell 100 Hue test. The 100-Hue test involves arranging 85 colored caps in a specific sequence based on their hue level. The caps are divided into 4 groups, with the first and last caps in each group fixed in place, serving as anchors. The test begins with all caps in random order, except for the fixed caps at the beginning and end of each group. The observer's task is to arrange the caps in the correct order according to color. This test measures an individual's ability to discern small differences in hue. 

\begin{equation} \label{eq:ColorTotalErrorScore}
\begin{split}
& E_{i,g} = |O_{i-1,g} - O_{i,g}| + |O_{i+1,g} - O_{i,g}| - 2\\
& \text{Total Error Score for group } g: \text{TES}_g = \sum_{i=2}^{n_g-1} E_{i,g} \\
& \text{Overall Total Error Score: TES} = \sum_{g=1}^{4} \text{TES}_g \\
& E_{i,g} \text{ is the error score of cap } i \text{ in group } g\\
& O_{i,g} \text{ is the order of the i-th cap in group } g\\
& n_g \text{ is the total number of caps in group } g \\
\end{split}
\end{equation}

The primary metric derived from the 100-Hue test is the Total Error Score (TES). Errors occur whenever caps are placed out of their correct sequence. The error score for a cap is calculated based on the distances between that cap and the caps immediately adjacent to it. Specifically, the score of a cap is the sum of the absolute differences between the number of that cap and the number of the caps on either side of it. For example, if cap number 10 is incorrectly placed between cap numbers 15 and 16, its score would be calculated as follows: \( |15 - 10| + |16 - 10| = 5 + 6 = 11 \). The error score for this cap is then derived by subtracting 2 from this sum, resulting in \( 11 - 2 = 9 \). If the cap had been correctly positioned, the score would have been calculated as \( |10 - 9| + |11 - 10| = 1 + 1 = 2 \), and after subtracting 2, the error score would be zero \cite{ghose2014simple}. TES for the entire set of caps is the sum of the error scores of all individual caps and quantifies the accuracy of color discrimination. The calculation is summarized in \Cref{eq:ColorTotalErrorScore}.

The 100-Hue test is particularly useful for detecting subtle color vision deficiencies that may not be apparent in more basic color vision tests. It is particularly relevant for professions where accurate color discrimination is critical. Normal values for the 100-Hue test vary depending on age and lighting conditions, but generally, a lower TES indicates better color discrimination ability. Approximately 16\% of the population makes 0 to 4 transpositions on the first test or has TES between 0 and 16, indicating superior color discrimination (Superior (good) score). About 68\% of the population scores between 16 and 100 on the first test, reflecting a normal range of color discrimination competence (Average (normal) score). Around 16\% of the population has a TES above 100, indicating poorer color discrimination (Low (weak) score). These ranges describe color-discrimination performance in the 100-Hue task; not reaching the superior range does not by itself indicate a clinical color-vision deficiency. Typically, the first retest may show some improvement, but further retests do not significantly affect the score \cite{Cwierz2021}.
\begin{figure}[!ht]
 \centering\includegraphics[width=0.85\columnwidth]{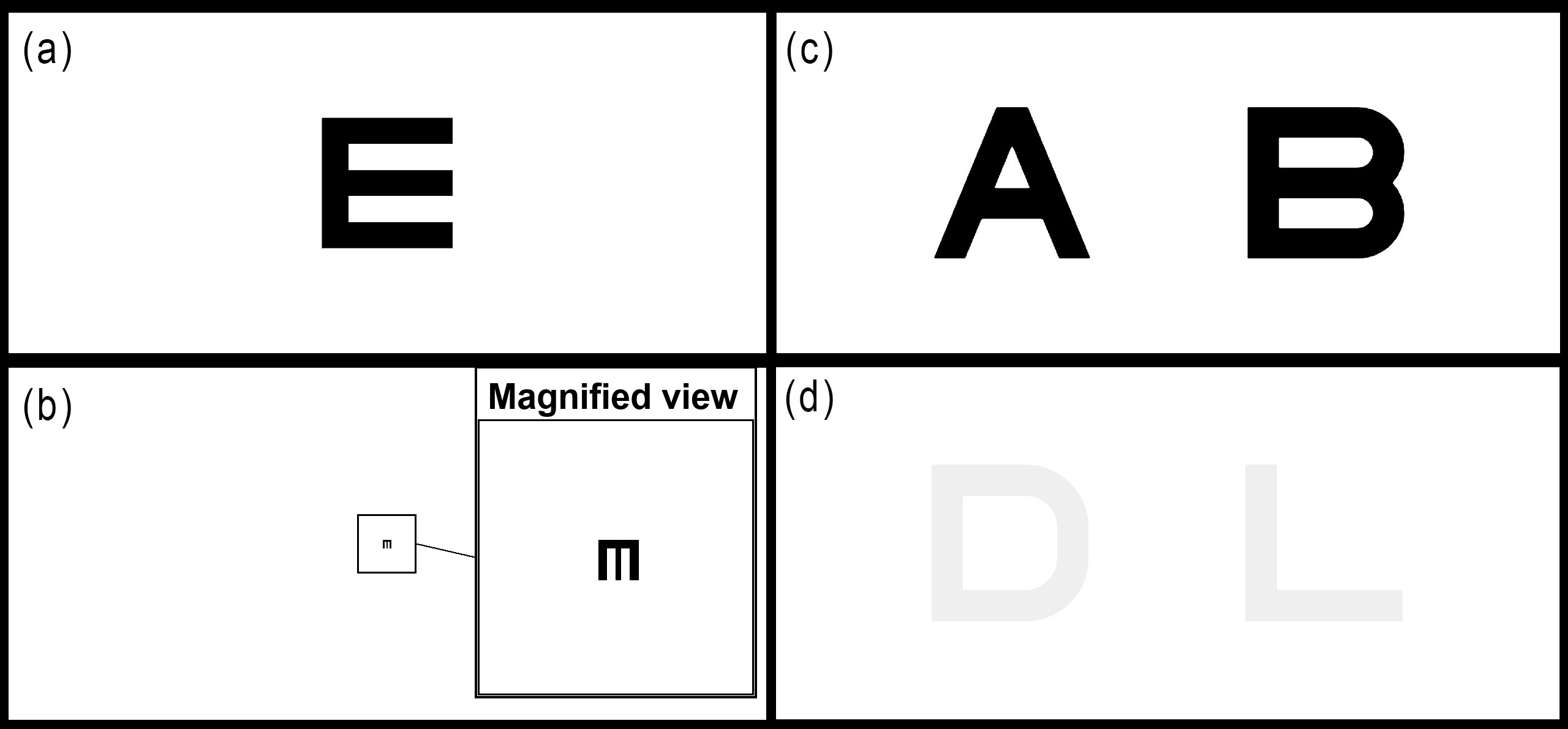}
 \caption{Screenshots of the visual acuity ((a) and (b)) and contrast sensitivity tests ((c) and (d)). The first and second rows ((a) (c), (b) (d)) are start and end state examples of the two tests. }
 \label{fig:CS_TEC}
\end{figure}

\begin{figure}[!ht]
 \centering\includegraphics[width=0.7\columnwidth]{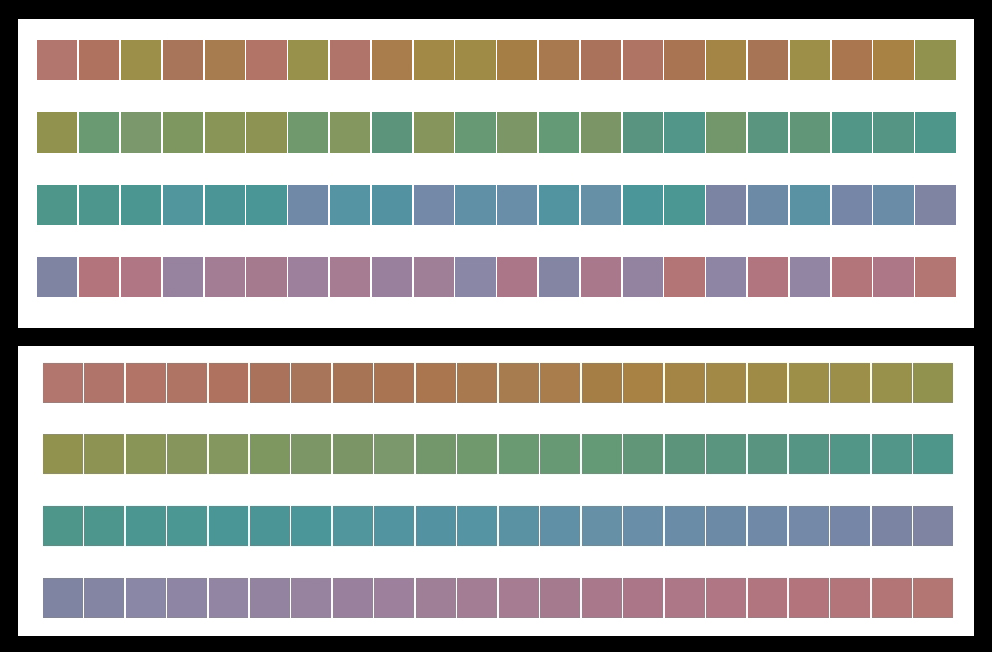}
 \caption{Screenshots of the digital version of the Farnsworth-Munsell 100 Hue test. The first and second rows are start and end state examples of the 100-Hue test. }
 \label{fig:100Hue}
\end{figure}

\begin{figure}[!ht]
 \centering\includegraphics[width=0.7\columnwidth]{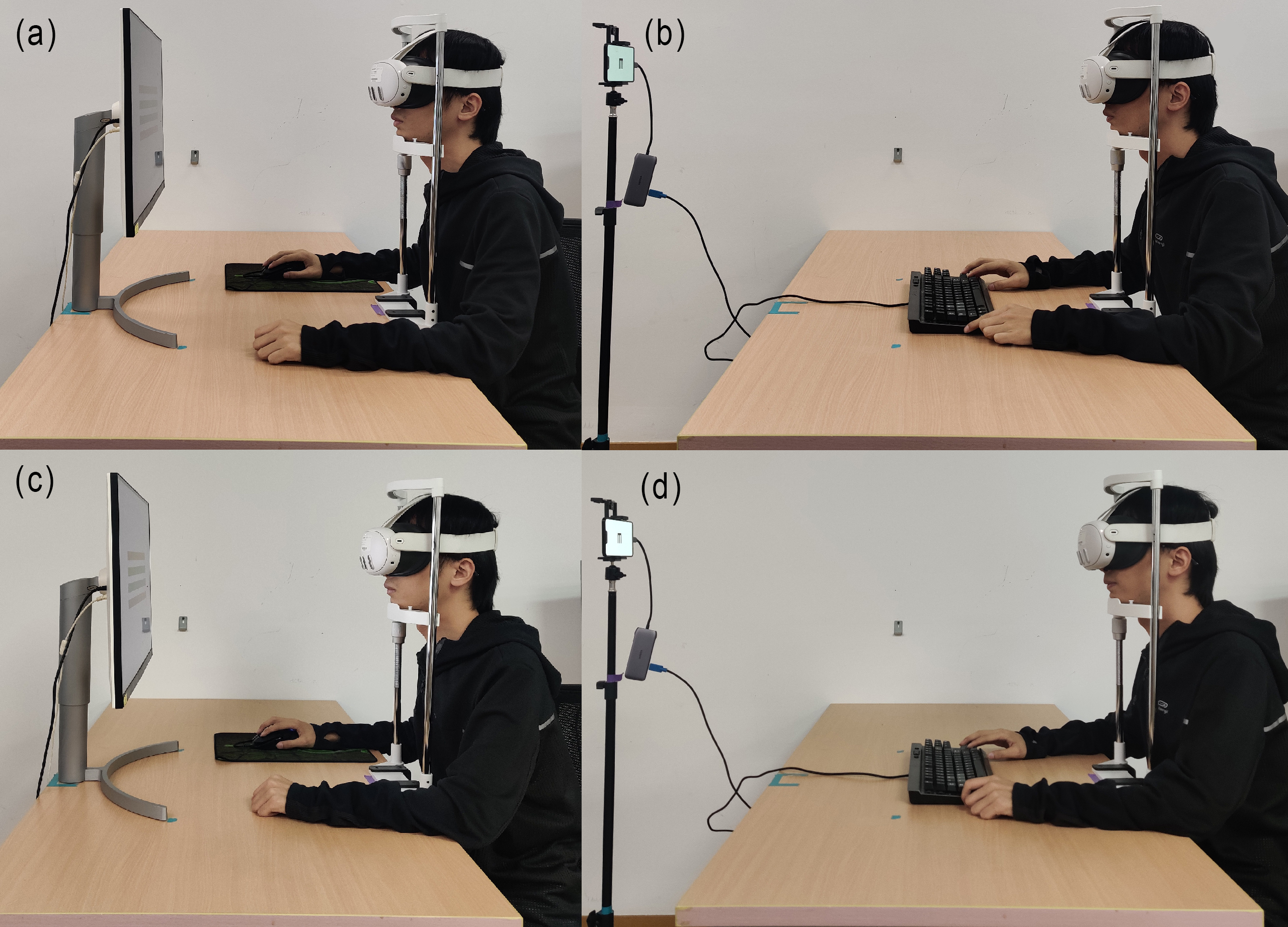}
 \caption{Pictures of the testing environment. (a) and (c): a 100-Hue test using the monitor. (b) and (d): tests using a smartphone. (a) and (b) were taken under the normal-light level (572 lux). (c) and (d) were taken under the low-light level (117 lux).}
 \label{fig:lightCondition}
\end{figure}

\begin{figure}[!ht]
 \centering\includegraphics[width=0.9\columnwidth]{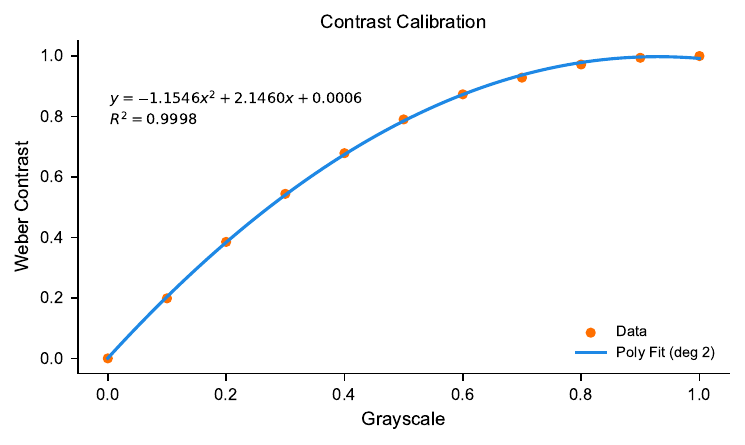}
 \caption{The polynomial fit of grayscale vs Weber contrast of Google Pixel 3 XL under 100\% screen brightness. }
 \label{fig:CS_Fit}
\end{figure}

\subsection{Program Design for the Digital Tests} \label{sec:programs}

For our \textit{visual acuity} test, we developed a smartphone-based tumbling E chart (TEC). Unlike traditional Snellen charts that use multiple letters or numbers, the TEC consists of multiple instances of the single capital letter "E" (an optotype) in different orientations (up, down, left, right). This design is particularly useful for testing the vision of individuals who may not be familiar with the alphabet, such as young children or people who are illiterate \cite{nottingham2011historical}. To enhance the accuracy and efficiency of optotype size adjustment, we employed a bisection method similar to that used in a recent study on visual acuity for VR HMDs \cite{ovva}. The bisection method is a numerical technique used to find the roots of a continuous function. For the \textit{contrast sensitivity} test, we developed a smartphone-based Pelli-Robson chart, also utilizing the bisection method for adjusting letter contrast. Finally, for the \textit{color vision} test, we adopted a design similar to existing digital 100-Hue tests \cite{Murphy2015}. Due to the large number of caps (85), we implemented this test using a PC program with a monitor to ensure accurate color representation and ease of use. The stimulus device was fixed by test type: the smartphone was used for visual acuity and contrast sensitivity, whereas the monitor was used only for color vision.

All programs were developed in Unity. For the TEC, we used a standard 5$\times$5 grid E optotype, displaying only one optotype in the center at a time to avoid the deviation caused by the spatial locations of multiple optotypes. Our recruited participants needed to answer the E letter orientation using the arrow keys of a keyboard connected to the phone via a USB hub. Eight continuous correct answers would reduce the E letter size, while wrong answers would trigger a fallback according to the bisection method. 

During pilot testing, we observed that participants made more errors when attempting to use the arrow keys to provide responses by themselves. The requirement of approximately 80 entries for each round caused participants to rush, increasing the likelihood of mistakes. To mitigate this, we adopted the traditional method where the researcher would enter the responses, ensuring greater accuracy and reducing the potential for input errors.

For the contrast sensitivity test, we used the Sloan font file created by Denis Pelli based on Louise Sloan’s specifications and used for the Pelli-Robson chart \footnote{\url{https://github.com/denispelli/Eye-Chart-Fonts}}. Participants needed to tell the two letters to a researcher, who would record the answers using a keyboard connected to the phone via a USB hub. Correct answers would reduce the grayscale of the letters, while wrong answers would trigger a fallback according to the bisection method. \Cref{fig:CS_TEC} shows the start and end state examples of the two tests. The TEC and contrast sensitivity test ended when the difference from the previous value was smaller than a default threshold (0.001 and 0.0001 for the two tests in our case). For the 100-Hue test, as shown in \Cref{fig:100Hue}, the caps within each line (with the first and final one in each line fixed) were sorted in a random order at the start of the test. Participants needed to rearrange the caps in hue order using a mouse. The test ended when they believed all the colors were in the correct order.

\subsection{Conditions and Procedure}\label{sec:conditions}
\Cref{fig:workflow} summarizes the stimulus presentation, VST viewing setup, lighting groups, measured visual metrics, and normalized performance summary used in the benchmark. To evaluate the visual perception performance, we utilized four testing conditions: Meta Quest 3, Meta Quest Pro, Apple Vision Pro (Quest 3, Quest Pro, Vision Pro for short), and without wearing a device (i.e., naked eyes). These three VR HMDs feature colorful RGB passthrough capabilities. Viewing condition was treated as a within-subject factor with these four levels, so each participant completed the naked-eye and three VST-HMD conditions. The order of the four testing conditions and the three visual perception tests was counterbalanced using a Latin square approach. Participants were divided into two equal groups to conduct the experiment under two different lighting conditions: normal light (with an illuminance of about 572 lux) and low light (about 117 lux), as shown in \Cref{fig:lightCondition}. To reduce fatigue and light-adaptation effects from repeating the full device-by-test protocol, lighting was treated as a between-subject factor, with each participant assigned to one lighting level. The illuminance levels were the average values measured using a DLX-LSK2304 illuminance meter (measuring range: 0 to 200,000 lux, with a precision of $\pm$4\% for values $\leq$ 10,000 lux) at eye level, with the lighting conditions set by adjusting the ceiling lights.

Prior to the formal testing, participants underwent a tutorial session to ensure familiarity with all visual perception tests. Rest periods were provided after each testing condition to prevent fatigue. The TEC and contrast sensitivity tests were untimed, allowing participants to complete them at their own pace. The 100-Hue test has a minimum time requirement of 8 minutes, based on the average completion time reported in previous research \cite{Murphy2015}. This duration, along with the tutorial session, would minimize the likelihood of needing a retest for the 100-Hue test \cite{Cwierz2021}. All tests measure two-eye vision. Each testing condition required approximately 12 minutes to complete, making the total duration for all four conditions around 1 hour.

\subsection{Participants and Apparatus}\label{sec:participants}
We recruited 24 participants (11 males and 13 females) with ages ranging from 19 to 38 ($M=22.63, SD=4.22$). The experiments were classified as low-risk research, adhered to the University's ethics guidelines and regulations, and received approval from its the University Ethics Committee at Xi'an Jiaotong-Liverpool University (\#ER-LRR-1288940720240411120606). All participants joined the experiment voluntarily and provided their consent. They had normal or corrected-to-normal eyesight and no self-reported color-vision deficiency during screening. The experiment was conducted in a closed-door lab with a constant temperature.

The sample size was checked with power analyses for the \(2 \times 4\) mixed design, with lighting as the between-subject factor and viewing condition as the within-subject factor. An a priori analysis indicated that 24 participants were sufficient for adequate power under a moderate effect size assumption (\(f=0.25\)), \(\alpha=0.05\), and desired power of 0.80; the achieved power for this sample size was 0.8157. A post hoc analysis using the observed large effects indicated power close to 1.0 for the main effects detected in the study.

\subsubsection{Calibrations}
The digital 100-Hue test was the only monitor-based test in our protocol. Weber-contrast calibration was only needed for the smartphone-based contrast sensitivity test, because the 100-Hue score is based on hue ordering rather than letter luminance contrast. Its accuracy depends on the performance of the monitor used to display the colors. A monitor with poor color accuracy can significantly affect the results of the test, as it may not accurately represent the subtle differences in hues that the test relies on \cite{Murphy2015}. To mitigate this issue, we used an LG 27UK650 monitor, a 4K in-plane switching (IPS) monitor that featured HDR10 (10-bit color depth that supported 1.07 billion colors) and covered 99\% of the standard RGB (sRGB) color gamut \footnote{\url{https://www.lg.com/us/monitors/lg-27uk650-w-4k-uhd-led-monitor}}. This setup provided a controlled display condition for comparing color-discrimination performance between VST HMDs and naked-eye viewing.

For the TEC and contrast sensitivity test, we used a Google Pixel 3 XL, an Android smartphone with a high resolution and contrast display. It had a 3K high resolution 2960$\times$1440 display with a pixels per inch (PPI) of 523 and used diamond sub-pixels with sub-pixel rendering to enhance sharpness and achieve higher peak brightness. The Pixel 3 XL display appears perfectly sharp for individuals with normal 20/20 vision at typical smartphone viewing distances of 25 to 46 cm. Due to its high brightness and low reflectance, the Pixel 3 XL also has a good contrast rating for high ambient light, ranging from 94 to 101 \footnote{\url{https://www.displaymate.com/Pixel_3XL_ShootOut_1g.htm}}. To ensure consistency and comparability with other studies, we calibrated the smartphone setup for the two smartphone-based tests. For both the TEC and contrast sensitivity test, the smartphone was mounted on a fixed stand, ensuring a horizontal distance of 1 meter between the device and the participant's chin rest.

The chin rest was used to stabilize the participant's head, with the smartphone positioned parallel to the eye level. For the 100-Hue test, we set a constant horizontal distance of 50 cm between the monitor and the participant's chin rest. No forehead rest was used to accommodate participants wearing VR HMDs. For the TEC, we measured the physical size of the E letter on the screen to determine the valid visual angle of the gap in the E letter. The screen brightness of the Google Pixel 3 XL was set to 100\% for all smartphone-based tests to maintain consistency. With the above configuration, our pilot testing indicated that the smallest integral E letter (without aliasing) displayed by the TEC corresponds to -0.62 in logMAR, where 0 represents normal vision. For the smartphone-based contrast sensitivity test, we used a screen luminance meter (SM208 luminance meter with a measuring range of 0.01 to 39,990 cd/m\(^2\) and a precision of $\pm$8\%) to measure the actual luminance of the letters on the Pixel 3 XL under different grayscale values and create the Weber contrast map shown in \Cref{fig:CS_Fit}. We used a polynomial equation to convert the grayscale value of the smartphone letters to Weber contrast for contrast sensitivity calculation. In VST conditions, logCS therefore denotes contrast sensitivity for the same calibrated physical letter stimulus as viewed through each HMD's passthrough pipeline, rather than isolated camera or display contrast.

\subsection{Results}\label{sec:results}
\begin{figure*}[!t]
 \centering\includegraphics[width=\textwidth]{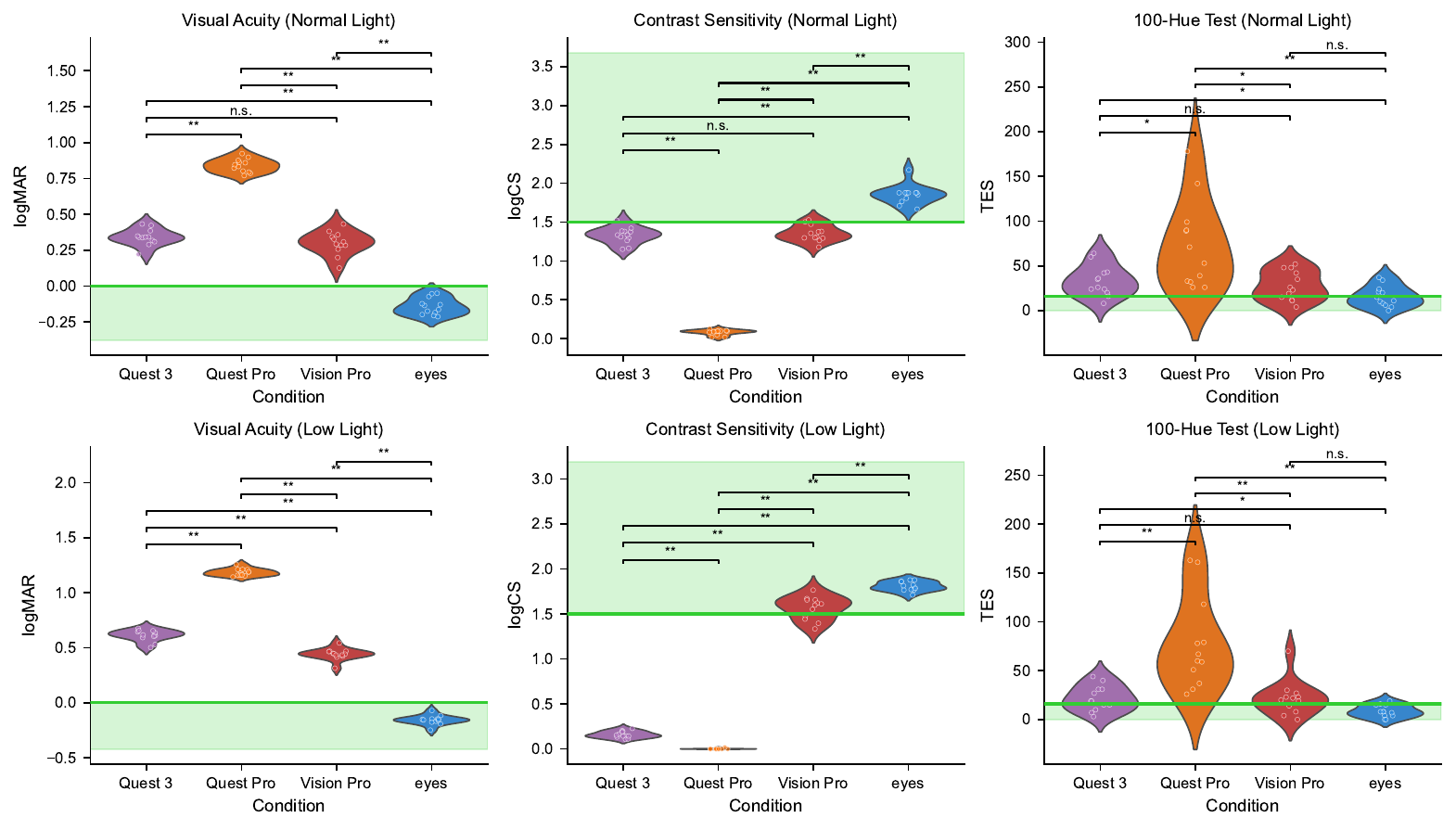}
 \caption{Violin plots with post-hoc results of the visual perception benchmark dataset. `*' to `***' represent Bonferroni-adjusted significant differences at `.05', `.01', `.001' level. The green area represents normal and above-normal logMAR and logCS values, and the superior TES range.}
 \label{fig:Violinplot}
\end{figure*}

\begin{figure*}[!t]
 \centering\includegraphics[width=\textwidth]{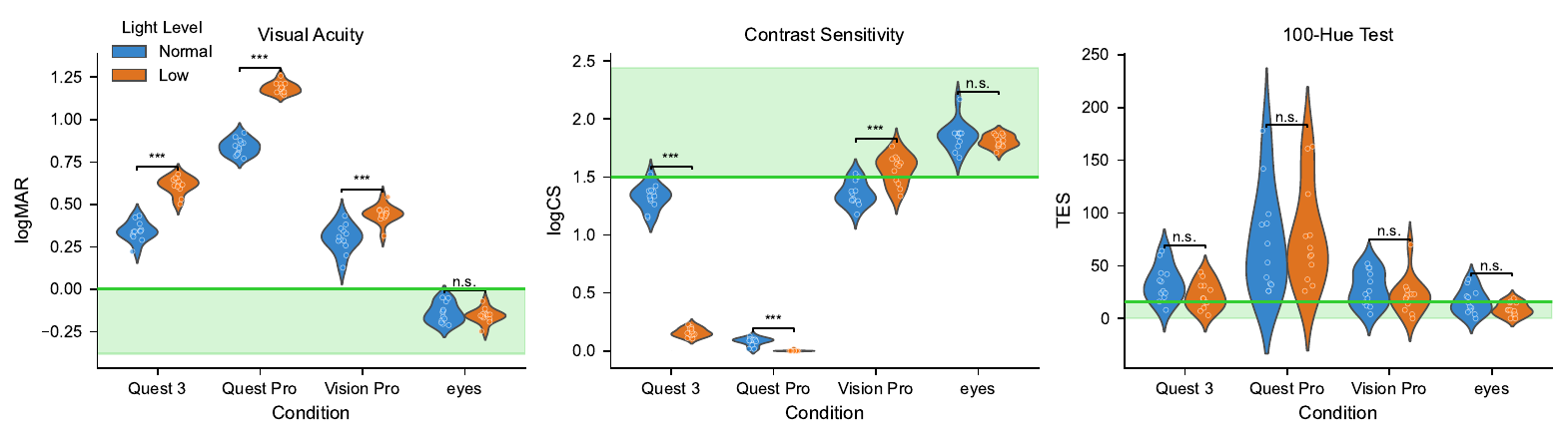}
 \caption{Violin plots for the normal vs. low-light level from the visual perception benchmark dataset. `*' to `***' represent significant differences at `.05', `.01', `.001' level. The green area represents normal and above-normal logMAR and logCS values, and the superior TES range.}
 \label{fig:AB_Violinplot}
\end{figure*}

\begin{figure}[!t]
 \centering\includegraphics[width=\columnwidth]{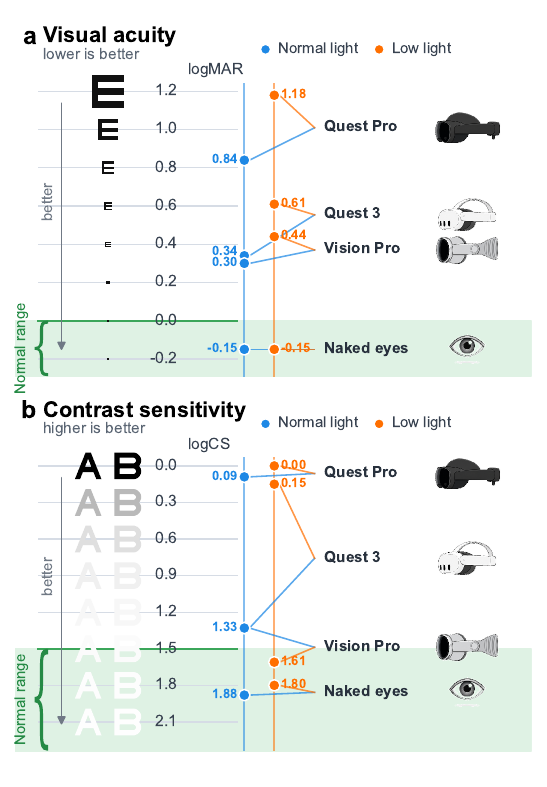}
 \caption{Reference-scale view of median visual acuity and contrast sensitivity. Blue and orange markers show normal- and low-light medians on separate axes for (a) logMAR and (b) logCS. Green braces and shading indicate the normal-reference ranges.}
 \label{fig:idealVST}
\end{figure}

\Cref{fig:idealVST} maps the normal-light and low-light median visual-acuity and contrast-sensitivity values of Quest Pro, Quest 3, Vision Pro, and naked-eye viewing onto the corresponding optotype scales used in the tests. We used non-parametric Friedman tests for each metric and lighting level because the data were non-normal. The Friedman tests showed significant differences across viewing conditions for all three visual functions in both lighting groups. We then used Wilcoxon signed-rank tests with Bonferroni correction for pairwise viewing-condition comparisons, with significant pairs summarized visually in \Cref{fig:Violinplot}. To evaluate lighting effects, we used Mann-Whitney U tests between the normal-light and low-light groups; these comparisons are summarized in \Cref{fig:AB_Violinplot}. The complete Friedman table with Kendall's $W$, the pairwise Wilcoxon table with effect-size $r$, and the Mann-Whitney U table are provided in the supplementary material available through the associated GitHub repository (\url{https://github.com/Chaosikaros/VST-Visual-Perception-Benchmark}), keeping the main paper focused on the interpretable summaries.

\subsection{Discussion}\label{sec:discussion}
 \Cref{fig:idealVST} shows the median acuity and contrast-sensitivity gaps against the normal-reference rows. All three HMDs have high logMAR values under both normal and low light levels, worse than the normal reference of human visual acuity limit: $\leq 0$. This degradation suggests that users may experience blurred or less detailed visuals, which can detract from the immersive experience that VST HMDs aim to provide.

\Cref{fig:Violinplot} shows that for logMAR and logCS, natural vision (eyes) significantly outperforms all three VST HMDs. Specifically, for logMAR and logCS in the normal-light level, there is no significant difference between Quest 3 and Vision Pro. However, in the low-light level, Vision Pro significantly outperforms Quest 3. None of these VST HMDs can ensure normal visual acuity (logMAR $\le 0$) in either light level. Regarding contrast sensitivity, Vision Pro is the only device that can ensure normal contrast sensitivity for some participants (logCS $\ge 1.5$). Quest 3 in the low-light level and Quest Pro in both light levels even exhibit a logCS that can be regarded as visual disability (logCS $\le 1.0$).

For color vision, some participants have poor color discrimination (TES $\ge 100$) with Quest Pro. Quest 3 and Vision Pro can ensure superior to average TES scores in normal and low-light levels. For TES, Quest Pro has significantly higher values than Quest 3 and Vision Pro in both normal and low-light levels. Additionally, Quest Pro and Quest 3 in both light levels have significantly higher TES values than natural vision.

\Cref{fig:AB_Violinplot} indicates that the normal and low-light levels cause no significant difference for natural vision in all metrics and for VST HMDs in TES. However, light levels cause significant differences for VST HMDs in logMAR and logCS. For all three VST HMDs, the low-light level results in significantly worse visual acuity. For Quest 3 and Quest Pro, the low-light level also results in significantly worse contrast sensitivity. Vision Pro instead shows higher contrast sensitivity in the low-light group, highlighting that low-light passthrough performance can vary substantially with device-specific image-processing pipelines.

Taken together, the three metrics point to different weaknesses in the VST pipeline. Visual-acuity loss mainly affects fine spatial detail such as small text, object edges, and distant features, and can arise from camera sampling, lens quality, reprojection, rendering, or display sampling. Contrast-sensitivity results are more dependent on lighting and device-specific processing, suggesting that exposure control, sensor noise, tone mapping, and dynamic range are important factors for low-light passthrough. Color-discrimination loss is less affected by lighting in our data and is more device-specific, suggesting a stronger role for camera color response, white balance, color correction, and display reproduction. For application design, these results mean that passthrough interfaces should not rely on small text, low-contrast boundaries, or color-only cues without enlargement, contrast enhancement, or redundant shape and luminance cues.

The mixed-source central PPD values in \Cref{tab:keyFactors} should therefore be interpreted cautiously. As also suggested by OVVA, nominal resolution or PPD alone is an incomplete clarity descriptor because PPD depends on the measurement method and viewing direction, while official values often report only central or peak PPD without a detailed protocol \cite{ovva}. In our results, measured visual acuity does not follow the PPD order in \Cref{tab:keyFactors} (Vision Pro $\sim$55, Quest Pro $\sim$30, Quest 3 18), further supporting the use of end-to-end visual tests such as OVVA and the present benchmark for HMD clarity comparison.

The device ordering is also informative when read together with \Cref{tab:keyFactors}. Quest Pro's weaker performance is qualitatively consistent with its single RGB camera plus depth-camera architecture, which may constrain the color passthrough path compared with binocular RGB passthrough. Quest 3 and Vision Pro both use binocular RGB camera architectures, but Vision Pro's better low-light acuity and contrast performance, together with its HDR display-side support and official low-light optimization feature \footnote{\url{https://support.apple.com/en-sg/120321}}, suggests a pipeline that preserves more usable detail in dim scenes. These links remain pipeline-level interpretations because the benchmark measures the combined camera-processing-display path.

\section{Design Implications}\label{sec:insights}
The Discussion above interprets the measured perceptual losses; this section translates those results into feasible recommendations for VST users, content developers, and HMD manufacturers:
\begin{itemize}
\item \textbf{L1}. Users should check passthrough performance in the target lighting and task context before relying on VST for visually demanding activities. In our data, low light worsened visual acuity for all three HMDs and worsened contrast sensitivity for Quest 3 and Quest Pro; these functions are also linked to mobility and driving-related visual performance in prior vision research \cite{KLEIN2003644, HALLEMANS2010547, wood2005standard}. Tasks that require small-text reading, distant edge recognition, or movement through dim scenes should therefore be evaluated under the intended illumination.
\item \textbf{L2}. VST applications should avoid assuming that the same visual design transfers equally across HMDs. Because our results show acuity and contrast losses across HMDs and a stronger color-discrimination loss for Quest Pro, passthrough interfaces should use larger text and targets, high luminance contrast, and redundant shape or brightness cues when color encodes state or risk. These choices align with prior work showing that HMD clarity, image processing, and color correction can affect perceived visual quality \cite{NeuralPassthrough, Carkeet2021, finlayson2015color, nam2010color}.
\item \textbf{L3}. HMD manufacturers should report hardware parameters together with standardized end-to-end perceptual measurements. Camera resolution, aperture, display characteristics, and PPD remain useful, but \Cref{tab:keyFactors} and our measured visual-acuity results show that mixed-source central PPD cannot explain the observed device ordering; OVVA similarly motivates user-centric visual-acuity measurement for HMD clarity \cite{ovva}. Reporting calibrated visual-acuity, contrast-sensitivity, and color-discrimination results across lighting levels would make VST device comparisons more actionable.
\end{itemize}

\section{Limitations and Future Work}
The main limitation is that we only tested two light levels. Although the current results suggest that low light levels can degrade users' visual perception performance, a wider spectrum of light levels may be helpful in investigating the exact degradation of visual perception performance. Because lighting was tested between subjects, these lighting effects should also be interpreted as group-level comparisons. Future work should validate them with a fully within-subject lighting design when participant burden and adaptation effects can be controlled. Screen-camera sampling interactions are another limitation: because the HMD cameras viewed physical smartphone or monitor displays, Moir\'e or aliasing artifacts may arise from the display subpixel structure, camera sampling, and viewing distance. Smartphone and monitor stimulus displays are also less precise than specialized optical test equipment. Here they support a low-cost, human-vision-based comparison of HMD performance under the same calibrated setup within each test. Future VST benchmarks should characterize these screen-camera sampling artifacts directly within each calibrated stimulus setup.
Another limitation is that our current method is designed for the central visual field rather than the entire visual field, which differs from omnidirectional measurement approaches such as OVVA that can assess the visual acuity distribution across the entire visual field in VR environments \cite{ovva}. However, to do this for VST would require a physical setup capable of including the whole visual field while maintaining consistent display performance. Such a setup can be quite expensive. Developing a practical low-cost setup for full-field VST measurement can be valuable in the future.

Furthermore, the three visual perception tasks, while representing the most commonly used tests, may not cover all aspects affecting the perception of the human eye. The 100-Hue task may also reflect spatial rendering and mouse-based cap ordering, so TES should not be interpreted as an isolated chromatic-fidelity measure. Future work will further separate these factors using dedicated VST setups. We also plan to measure the differences in stereo vision and depth perception. 
To ensure a safe and feasible user experience, current VST HMDs could include additional vision enhancement features with environmental sensors to compensate for the difference between common camera systems and human eyes. Vision-chip-based approaches for robust visual perception may also inform future VST sensing pipelines \cite{yang2024vision}. While OST HMDs do not face similar challenges in certain aspects, such as achieving normal visual acuity in the background view, VST HMDs have unique advantages that should not be compromised by the degradation of visual perception performance.

\section{Conclusion}
We evaluated visual acuity, contrast sensitivity, and color vision for Vision Pro, Quest 3, and Quest Pro under normal and low light, using naked-eye viewing as the reference. Natural vision outperformed all three HMDs across the tasks, and no device matched normal visual acuity at either light level. For contrast sensitivity, Quest 3 and Vision Pro performed similarly under normal light, but Vision Pro outperformed Quest 3 under low light; Quest Pro performed worst overall; only vision Pro reached the normal-reference range for some participants under low light. Lighting effects were device- and metric-specific. Low light reduced visual acuity for all HMDs and contrast sensitivity for Quest 3 and Quest Pro. Vision Pro instead showed higher contrast sensitivity in the low-light group. Color-discrimination performance varied more by device than by lighting. Overall, naked-eye viewing nevertheless remained significantly better.

Because the benchmark measures the combined camera-processing-display pipeline, the observed losses cannot be attributed to a single component. Nonetheless, the results provide an end-to-end basis for comparing devices and identifying conditions that may require interface compensation. This distinction matters for device selection and interface design in tasks that depend on fine detail, low contrast, or reliable color discrimination. The results also show why VST performance should be tested in the intended lighting and task context rather than inferred from individual hardware specifications. Future work should cover more light levels and extend the benchmark to additional visual functions and full-field viewing.

\acknowledgments{
We thank the participants for their time and the reviewers for their insightful comments and helpful suggestions. Part of this research was conducted while Jialin Wang was affiliated with Xi’an Jiaotong-Liverpool University.}

\bibliographystyle{abbrv-doi}

\bibliography{MyCollection}
% Supplementary material appended after the unchanged CR manuscript.
\clearpage
\onecolumn
\section*{Supplementary Material}
\setcounter{table}{0}
\renewcommand{\thetable}{S\arabic{table}}
\setlength{\tabcolsep}{5pt}
\renewcommand{\arraystretch}{1.12}

\noindent\textbf{Repository:} \url{https://github.com/Chaosikaros/VST-Visual-Perception-Benchmark}

\section*{Supplementary Statistical Tables}

This supplementary material provides the detailed statistical tables moved from the main manuscript for readability. Q3, QP, VP, and eyes denote Quest 3, Quest Pro, Vision Pro, and naked-eye viewing, respectively. P-values are Bonferroni-adjusted where noted; highlighted cells indicate significant comparisons. For the Wilcoxon post-hoc table, \(W\) denotes the Wilcoxon signed-rank statistic, and \(r=|z|/\sqrt{n}\) is computed from the normal approximation using nonzero paired differences.

\begin{table}[H]
\centering
\small
\caption{Complete Friedman tests for the four viewing conditions: Quest 3, Quest Pro, Vision Pro, and naked-eye viewing.}
\label{tab:supp_friedman}
\begin{tabular}{lccccc}
\toprule
\textbf{Light Level} & \textbf{Metric} & \textbf{$\chi^2$} & \textbf{df} & \textbf{p} & \textbf{Kendall's $W$} \\
\midrule
\multirow{3}{*}{Normal} & logMAR & 34.6 & 3 & \highlightcell{$<$0.001} & 0.786 \\
 & logCS & 32.5 & 3 & \highlightcell{$<$0.001} & 0.739 \\
 & TES & 22.0 & 3 & \highlightcell{$<$0.001} & 0.500 \\
\midrule
\multirow{3}{*}{Low} & logMAR & 36.0 & 3 & \highlightcell{$<$0.001} & 0.818 \\
 & logCS & 36.0 & 3 & \highlightcell{$<$0.001} & 0.818 \\
 & TES & 26.6 & 3 & \highlightcell{$<$0.001} & 0.604 \\
\bottomrule
\end{tabular}
\end{table}

\begin{table}[H]
\centering
\scriptsize
\setlength{\tabcolsep}{3.5pt}
\caption{Complete Wilcoxon signed-rank post-hoc tests with Bonferroni correction and effect-size \(r\) for pairwise viewing-condition comparisons.}
\label{tab:supp_posthoc}
\begin{tabular}{llrrrrrr}
\toprule
 & & \multicolumn{3}{c}{\textbf{Normal light}} & \multicolumn{3}{c}{\textbf{Low light}} \\
\cmidrule(lr){3-5} \cmidrule(lr){6-8}
\textbf{Metric} & \textbf{Pair} & \textbf{W} & \textbf{r} & \textbf{p} & \textbf{W} & \textbf{r} & \textbf{p} \\
\midrule
\multirow{6}{*}{logMAR} & Q3 vs. QP & 0 & 0.88 & \highlightcell{0.003**} & 0 & 0.88 & \highlightcell{0.003**} \\
 & Q3 vs. VP & 3 & 0.79 & 0.059 & 0 & 0.88 & \highlightcell{0.003**} \\
 & Q3 vs. eyes & 0 & 0.88 & \highlightcell{0.003**} & 0 & 0.88 & \highlightcell{0.003**} \\
 & QP vs. VP & 0 & 0.88 & \highlightcell{0.003**} & 0 & 0.88 & \highlightcell{0.003**} \\
 & QP vs. eyes & 0 & 0.88 & \highlightcell{0.003**} & 0 & 0.88 & \highlightcell{0.003**} \\
 & VP vs. eyes & 0 & 0.88 & \highlightcell{0.003**} & 0 & 0.88 & \highlightcell{0.003**} \\
\midrule
\multirow{6}{*}{logCS} & Q3 vs. QP & 0 & 0.88 & \highlightcell{0.003**} & 0 & 0.88 & \highlightcell{0.003**} \\
 & Q3 vs. VP & 32 & 0.16 & 1.000 & 0 & 0.88 & \highlightcell{0.003**} \\
 & Q3 vs. eyes & 0 & 0.88 & \highlightcell{0.003**} & 0 & 0.88 & \highlightcell{0.003**} \\
 & QP vs. VP & 0 & 0.88 & \highlightcell{0.003**} & 0 & 0.88 & \highlightcell{0.003**} \\
 & QP vs. eyes & 0 & 0.88 & \highlightcell{0.003**} & 0 & 0.88 & \highlightcell{0.003**} \\
 & VP vs. eyes & 0 & 0.88 & \highlightcell{0.003**} & 0 & 0.88 & \highlightcell{0.003**} \\
\midrule
\multirow{6}{*}{TES} & Q3 vs. QP & 4.5 & 0.78 & \highlightcell{0.023*} & 0 & 0.89 & \highlightcell{0.006**} \\
 & Q3 vs. VP & 24 & 0.34 & 1.000 & 35.5 & 0.08 & 1.000 \\
 & Q3 vs. eyes & 6.5 & 0.74 & \highlightcell{0.044*} & 4 & 0.79 & \highlightcell{0.018*} \\
 & QP vs. VP & 2 & 0.83 & \highlightcell{0.018*} & 0 & 0.88 & \highlightcell{0.003**} \\
 & QP vs. eyes & 0 & 0.88 & \highlightcell{0.003**} & 0 & 0.88 & \highlightcell{0.003**} \\
 & VP vs. eyes & 5.5 & 0.74 & 0.064 & 3 & 0.79 & 0.059 \\
\bottomrule
\end{tabular}
\end{table}

\begin{table}[H]
\centering
\small
\caption{Complete Mann-Whitney U tests comparing normal-light and low-light groups for each metric and viewing condition.}
\label{tab:supp_mwu}
\begin{tabular}{llrr}
\toprule
\textbf{Metric} & \textbf{Condition} & \textbf{U} & \textbf{p} \\
\midrule
\multirow{4}{*}{logMAR} & Quest 3 & 0 & \highlightcell{$<$0.001} \\
 & Quest Pro & 0 & \highlightcell{$<$0.001} \\
 & Vision Pro & 7 & \highlightcell{$<$0.001} \\
 & eyes & 77 & 0.795 \\
\midrule
\multirow{4}{*}{logCS} & Quest 3 & 144 & \highlightcell{$<$0.001} \\
 & Quest Pro & 142 & \highlightcell{$<$0.001} \\
 & Vision Pro & 14.5 & \highlightcell{$<$0.001} \\
 & eyes & 94.5 & 0.195 \\
\midrule
\multirow{4}{*}{TES} & Quest 3 & 101 & 0.10 \\
 & Quest Pro & 68 & 0.84 \\
 & Vision Pro & 88 & 0.37 \\
 & eyes & 99.5 & 0.117 \\
\bottomrule
\end{tabular}
\end{table}
\end{document}